\documentclass{article}
\usepackage{mathrsfs}
\usepackage{mathtools}
\usepackage{comment} 
\usepackage[margin=3.5cm]{geometry}
\usepackage{booktabs, tabularx} 
\usepackage{enumerate}

\newcommand{\vF}{\overline{\F}_q}

\input{head}
\fxsetup{final}
\FXRegisterAuthor{a}{eh}{\color{cyan}Anuvrat} 
\FXRegisterAuthor{d}{ej}{\color{brown}Divyesh}  
\fxusetheme{color}

\title{Good rational functions up to degree six}
\author{
Anuvrat Jaindungarwal\thanks{Department of Mathematics and Statistics, University of South Florida, Tampa, FL 33620, USA. Email: \href{mailto:anuvrat@usf.edu}{\nolinkurl{anuvrat@usf.edu}}.}
\and
Divyesh Vaghasiya\thanks{Department of Mathematics and Statistics, University of South Florida, Tampa, FL 33620, USA. Email: \href{mailto:divyeshvaghasiya00@gmail.com}{\nolinkurl{divyeshvaghasiya00@gmail.com}}.}
}
\date{}

\begin{document}

\maketitle
\begin{abstract}
    Locally recoverable codes (LRCs) are fundamental to the architecture of modern distributed storage systems. The Tamo-Barg construction of optimal LRCs was revolutionized by the introduction of good polynomials, which group evaluation points into local recovery sets. However, polynomial constructions are constrained by their inherently large Galois groups, which limit the number of totally split rational places. Recent advances have generalized this paradigm to good rational functions, which leverage the geometry of the projective line to achieve minimal Galois groups and subsequently codes with greater length. In this paper, we complete the classification of good rational functions of degree up to $6$ in characteristic $2$. By embedding the problem into algebraic function field theory, we explicitly determine the minimal arithmetic and geometric Galois groups for these low degrees. Furthermore, we move beyond asymptotic Hasse-Weil estimates by proving that the exact number of totally split places over extension fields obeys a linear recurring sequence (LRS). This gives an efficient method for computing the exact lengths and dimensions of the corresponding optimal LRCs and, in many cases, yields parameters beyond those obtainable from good polynomials.
\end{abstract}

\section{Introduction}

Let $q$ be a prime power. An $(n,k,r)$-locally recoverable code (LRC) $\mathcal{C}$ is a $k$-dimensional linear subspace of $\F_q^n$ characterized by the property that any single erased coordinate of a codeword can be reconstructed by accessing at most $r$ other coordinates. To maximize storage efficiency, constructions aim to hit the Singleton-type upper bound on the minimum distance established by Gopalan et al.~\cite{Gopalan2012}:
$$d(\mathcal{C}) \le n - k - \left\lceil \frac{k}{r} \right\rceil + 2$$
Codes achieving this bound are called optimal. The foundational mechanism for constructing optimal LRCs relies on the Tamo-Barg evaluation framework ~\cite{TamoBarg2014}, which uses ``good polynomials'' of degree $r+1$ that remain constant on $\ell$ disjoint subsets of $\F_q$ of size $r+1$. 

As established by Micheli ~\cite{Micheli}, finding optimal parameters for these codes over a fixed field translates directly into a Galois-theoretic problem over global function fields. The number of disjoint local recovery sets $\ell$ corresponds to the number of totally split rational places $\#T_{\mathrm{split}}^1(f)$ in the global function field extension $\F_q(x)/\F_q(t)$ defined by $f(x)=t$. Consequently, maximizing $\ell$ requires finding polynomials that minimize the size of the arithmetic Galois group $A(f)\defn \Gal(f(x)-t / \F_q(t))$.

\subsection{Good Polynomials and Exact Counting}
This Galois-theoretic perspective has led to extensive efforts to classify low-degree polynomials with minimal monodromy groups. Dukes, Ferraguti, and Micheli determined the minimum possible
orders of these groups for polynomials of degree at most $5$ over finite
fields and provided explicit estimates for the corresponding numbers of
totally split rational places \cite{Giacomo}. Recently in \cite{good_poly_self}, the low-degree polynomial classification paradigm was completed up to degree $6$. A core focus of that work was mapping the difficult scenarios where the field characteristic divides the degree of the polynomial, which are both highly technical to resolve algebraically and highly relevant for practical cryptographic and code implementations.

Beyond group classification, a critical milestone in the polynomial framework was the introduction of an exact counting methodology over extension fields. While standard evaluations rely heavily on the Hasse--Weil bound to provide asymptotic estimates for number of totally split rational places, they fail to provide exact parameters when the base field is extended. To address this, it was proven
in \cite[Theorem~4.5]{good_poly_self} that the sequence
\[
a_i=\#\left\{t_0\in\F_{q^i}:
f(X)-t_0\text{ splits into }\deg(f)\text{ distinct linear factors over }
\F_{q^i}\right\}
\]
is a linear recurring sequence over $\Z$. This allowed the exact maximum length and dimension of a Tamo-Barg code to be computed via a predictable recurrence relation rather than brute-force counting.

\subsection{From Good Polynomials to Good Rational Functions}

Despite the precision of the polynomial framework, it suffers from an intrinsic algebraic bottleneck. The Galois group of a generic polynomial of degree $r+1$ is bounded by the symmetric group $S_{r+1}$, which restricts the density of totally split places to roughly $q/(r+1)!$. Moreover, in several low-degree cases, the smallest Galois group attainable
by a polynomial has order strictly greater than $r+1$. To overcome this limitation, transitioning from polynomials to good rational functions $h(x) = f(x)/g(x) \in \F_q(x)$ was recently suggested as the primary avenue for future research in \cite{good_poly_self} to bypass these structural bounds. Concurrently, recent independent work has begun to formalize this transition in~\cite{Beyond_polynomials}. Because rational functions take values on the projective line $\mathbb{P}^1(\vF) = \vF \cup \{\infty\}$, they introduce a more flexible ramification structure. We note that rational maps have previously appeared in LRC constructions, such as in the work of Munuera and Ten\'orio \cite{MUNUERA}.

Formally, let $h=f/g\in\F_q(x)$ be a separable rational function, where $f,g\in\F_q[X]$ are coprime. Let $M_h$ denote the splitting field of $f(X)-tg(X)$ over $\F_q(t)$, and let $\overline{M}_h$ denote its splitting field over $\overline{\F}_q(t)$. We define the arithmetic and geometric Galois groups of $h$ by
\[
A(h)\defn\Gal(M_h/\F_q(t))
\qquad\text{and}\qquad
G(h)\defn\Gal(\overline{M}_h/\overline{\F}_q(t)),
\]
respectively. For every positive integer $n$, let $\cG_n(q)$ be the least positive integer such that there exists a rational function $h \in \F_q(x)$ of degree $n$ achieving minimal Galois groups of size $|A(h)|=|G(h)|=\cG_n(q)$.

A prominent framework developed by Liu, Mesnager, Tang, and Zheng systematically extended Micheli's Galois closure framework to rational functions. They proved that the extremal case occurs when the function field extension $\F_q(x)/\F_q(t)$ is itself Galois, collapsing the Galois group to its absolute minimal size, $\#G_h = r+1$. While they provided specific cyclic constructions for odd characteristics including a degree-3 construction for characteristic 3 ($q=3^m$) and tame constructions for $q \equiv 1 \pmod 3$ or $q \equiv 3 \pmod 4$. The systematic classification of low-degree rational functions in characteristic 2 remains an open problem.

\subsection{Equivalence under Fractional Linear Transformations}
When determining the minimal Galois groups $\mathcal{G}_n(q)$, it is sufficient to analyze rational functions up to equivalence under fractional linear transformations. More precisely, two rational functions $h_1, h_2 \in \F_q(x)$ are said to be equivalent if$$h_2 = \mu_1 \circ h_1 \circ \mu_2$$
for some $\mu_1, \mu_2 \in \mathrm{PGL}_2(\F_q)$. Postcomposition by $\mu_1$ does not change the corresponding subfield, since
  $$\F_q(\mu_1 \circ h_1) = \F_q(h_1),$$
whereas precomposition by $\mu_2$ induces an $\F_q$-automorphism of $\F_q(x)$. Therefore, such transformations strictly preserve the sizes of both the arithmetic and geometric Galois groups, $A(h)$ and $G(h)$. Let $\ell=\deg(h)$. By applying an appropriate transformation, we can always assume without loss of generality that $\operatorname{deg}f \le \ell-1$ and $\operatorname{deg}g = \ell$, where $g$ is monic. If, moreover, $\operatorname{char}(\F_q) \nmid \ell$, a translation of the variable allows us to assume that the coefficient of $x^{\ell-1}$ in $g$ is zero. 

\subsection{Our Contributions}
The recent framework of Liu et al \cite{Beyond_polynomials} establishes the connection between good rational functions, totally split rational places, and optimal LRCs. It also provides explicit constructions, particularly when the extension $\mathbb{F}_q(x)/\mathbb{F}_q(h(x))$ is Galois. However, the systematic classification of low-degree rational functions according to their arithmetic and geometric Galois groups remains open. In fact, such a classification is explicitly proposed as a direction for further research in their work.  Furthermore, the exact counting method developed for good polynomials has not previously been extended to rational functions. We address these two problems by directly bridging the methodology of low-degree polynomial classification with the newly established rational function paradigm.  Our specific contributions are organized as follows:
\begin{itemize}
    \item Exhaustive Characteristic $2$ Classification: We determine the minimum possible common order $\cG_n(q) = \abs{A(h)} = \abs{G(h)}$ for the remaining cases of rational functions of degrees at most $6$ over finite fields of characteristic $2$. Together with the cases already realized by polynomials, this completes the low-degree classification. We provide explicit rational functions attaining these absolute minimum values and derive bounds on the corresponding numbers of totally split rational places.
    \item  Porting the LRS Framework: We extend the linear recurring sequence (LRS) method from polynomials to rational functions. We prove that the exact number of totally split places for a rational function $f(X) - t_0 g(X)$ over extension fields $\mathbb{F}_{q^i}$ is governed by an LRS. This provides coding theorists with an exact and efficient method for computing the parameters of the associated optimal LRCs over extension fields, entirely bypassing the $\mathcal{O}(\sqrt{q})$ ambiguity left by asymptotic estimates.
\end{itemize}
 
\section{Preliminaries}
Throughout this paper, $q$ denotes a prime power, $\F_q$ denotes the finite field of order $q$, and $\mathbb{P}^1(\F_q) = \F_q \cup \{\infty\}$ denotes the projective line over $\F_q$. In this section, we recall the background on algebraic function fields, ramification, splitting of places, genus bounds, and linear recurring sequences that will be used throughout the paper.

\subsection{Algebraic Function Fields and Genus Bounds}
Let $E/F$ be a finite separable extension of algebraic function fields having the same full constant field $\F_q$. We denote their genera by $g_E$ and $g_F$, respectively. The Hurwitz genus formula gives
$$2g_E-2 = [E:F](2g_F-2) + \deg\bigl(\operatorname{Diff}(E/F)\bigr), \label{eq:hurwitz-different}$$ 
where $\operatorname{Diff}(E/F)$ denotes the different of the extension $E/F$~\cite[Theorem~3.4.13]{Stichtenoth}. 
For every place $P$ of $F$ and every place $Q$ of $E$ lying above $P$, let $e(Q\mid P)$ denote the corresponding ramification index. Since$$d(Q\mid P)\geq e(Q\mid P)-1,$$with equality if and only if $Q\mid P$ is tamely ramified, we obtain
\begin{equation}
    2g_E-2 \geq [E:F](2g_F-2) + \sum_{P\in\cP_F} \sum_{Q\mid P} \bigl(e(Q\mid P)-1\bigr)\deg(Q). 
\label{eq:hurwitz-ramification}
\end{equation}
Equality holds in \eqref{eq:hurwitz-ramification} if and only if the extension $E/F$ is tame~\cite[Corollary~3.5.6]{Stichtenoth}.We will also use Castelnuovo's inequality to bound the genera of splitting fields obtained as composita. Let $M/\F_q$ be an algebraic function field containing subfields $F_1$ and $F_2$ such that $M=F_1F_2$. Setting
$$n_i=[M:F_i] \qquad\text{and}\qquad g_i=g_{F_i}$$
for $i\in\{1,2\}$, Castelnuovo's inequality yields
\begin{equation}
    g_M \leq n_1g_1+n_2g_2+(n_1-1)(n_2-1).
\label{eq:castelnuovo}
\end{equation}

See~\cite[Theorem~3.11.3]{Stichtenoth}.

\subsection{Places, Ramification, and Splitting}

For an algebraic function field $F/\F_q$, let $\cP_F$ denote the set of
places of $F$, and let $\cP_F^1$ denote the set of its rational places.
Let $E/F$ be a finite separable extension, let $P\in\cP_F$, and let
$Q\in\cP_E$ be a place lying above $P$. We denote by
$e(Q\mid P)$ and $f(Q\mid P)$ the ramification index and the relative
degree of $Q$ over $P$, respectively. The fundamental equality states
that
\begin{equation}
    \sum_{Q\mid P}e(Q\mid P)f(Q\mid P)=[E:F].
    \label{eq:fundamental-equality}
\end{equation}

The place $P$ is ramified in $E/F$ if $e(Q\mid P)>1$ for some
$Q\mid P$. It is said to split completely in $E/F$ if there are exactly
$[E:F]$ distinct places of $E$ lying above $P$. Equivalently, 
$e(Q\mid P)=f(Q\mid P)=1$
for every $Q\mid P$; see~\cite[Definition~3.1.13]{Stichtenoth}.

Let
\[
h(x)=\frac{f(x)}{g(x)}\in\F_q(x),
\]
where $f,g\in\F_q[X]$ are coprime, and suppose that $h$ is nonconstant
and separable. Set $t=h(x)$ and $n=\deg(h)=\max\{\deg(f),\deg(g)\}$. 
Then $[\F_q(x):\F_q(t)]=n$, and the extension is described by
$f(X)-tg(X)=0$.

For $t_0\in\mathbb{P}^1(\F_q)$, let $P_{t_0}$ denote the corresponding
rational place of $\F_q(t)$. The fiber $h^{-1}(t_0)$ contains exactly
$n$ distinct points of $\mathbb{P}^1(\F_q)$ if and only if $P_{t_0}$
splits completely in $\F_q(x)/\F_q(t)$~\cite[Proposition~4.1]{Beyond_polynomials}. 
We therefore define
\[
T_{\mathrm{split}}^1(h)
\defn
\left\{
t_0\in\mathbb{P}^1(\F_q):
P_{t_0}\text{ splits completely in }\F_q(x)/\F_q(t)
\right\}.
\]

A rational function $h$ is said to be $(r,\ell)$-good if
$\deg(h)=r+1$ and there exist $\ell$ distinct points
$t_1,\ldots,t_\ell\in\mathbb{P}^1(\F_q)$ such that
\[
\#h^{-1}(t_i)=r+1
\]
for every $i\in\{1,\ldots,\ell\}$. Thus, the largest possible value of
$\ell$ for a fixed $h$ is
\[
\ell=\#T_{\mathrm{split}}^1(h).
\]
Each such fiber forms a local group of size $r+1$ in the associated LRC
construction; after deleting the erased coordinate, the remaining $r$
coordinates form its recovering set.

\subsection{Linear Recurring Sequences}

A sequence $(a_i)_{i\geq 1}$ of integers is called a linear recurring
sequence if there exist an integer $m\geq 1$ and coefficients
$c_1,\ldots,c_m\in\Z$ such that
\begin{equation}
    a_i= c_1a_{i-1} + c_2a_{i-2}+ \cdots + c_ma_{i-m}
    \label{eq:LRS}
\end{equation}
for every $i \geq m+1$.

Let $F/\F_q$ be an algebraic function field of genus $g$, and let
$
F_i=F\F_{q^i}
$
denote its constant field extension of degree $i$. If
$
N_i=\#\cP_{F_i}^1
$
is the number of rational places of $F_i/\F_{q^i}$, then
\begin{equation}
    N_i
    =
    q^i+1-\sum_{j=1}^{2g}\alpha_j^i,
    \label{eq:rational-place-count}
\end{equation}
where $\alpha_1,\ldots,\alpha_{2g}$ are the reciprocal roots of the
$L$-polynomial of $F/\F_q$; see
\cite[Corollary~5.1.16]{Stichtenoth}. Consequently, $(N_i)_{i\geq 1}$
is a linear recurring sequence over $\Z$.

In~\cref{sec:LRS}, we combine this fact with a separate analysis
of the ramified rational places to prove that the number of totally
split rational places associated with a rational function also forms a
linear recurring sequence. This extends the corresponding result for
polynomials obtained in~\cite{good_poly_self}.

\section{Minimal Galois Groups in Characteristic 2}

The primary obstacle in constructing optimal rational-function LRCs over fields of characteristic 2 is identifying functions that achieve the absolute minimal Galois group size. In this section, we systematically classify these minimal groups $\cG_n(q)$ for degrees up to 6. Our strategy proceeds in three steps. First, we establish the structural constraints on prime cyclic extensions using the Riemann-Hurwitz formula. Second, we bound the degree of the function field extension using the automorphism group of $\F_q(x)$. Finally, we apply these theoretical constraints to conduct an exhaustive computational search, isolating the explicit rational functions that achieve the theoretical minimums.

We begin by analyzing the ramification behavior of rational functions that induce cyclic Galois groups of prime order. Because we are ultimately interested in bounding the number of totally split places, understanding the factorization forced by tame ramification provides a necessary structural template for our search space.

\begin{Prop}
\label{prop:cyclic-prime}
Let $\ell$ be a prime with $\ell\nmid q$, and let $h\in\F_q(x)$ be a rational function of degree $\ell$ such that $A(h)=G(h)\cong C_\ell$. Up to postcomposition by an element of $\operatorname{PGL}_2(\F_q)$, write $h=f/g$, where $f,g\in\F_q[X]$ are coprime, $\deg f<\deg g=\ell$, and $g$ is monic. Then $\deg f\in\{0,\ell-1\}$. Moreover:
\begin{enumerate}[(i)]
    \item if $\deg f=0$, then $h$ is equivalent to a polynomial of degree $\ell$;
    \item if $\deg f=\ell-1$, then there exist $a,b\in\F_{q^2}$, with $b\neq0$, such that
    \[
    f(X)-bg(X)=-b(X-a)^\ell.
    \]
\end{enumerate}
\end{Prop}

\begin{proof}
Set $K=\F_q(t)$ and $L=\F_q(x)$, where $t=h(x)$. Since $A(h)=G(h)\cong C_\ell$, the splitting field of $f(X)-tg(X)$ over $K$ has degree $\ell$. As it contains $L$ and $[L:K]=\ell$, we conclude that $L/K$ is a cyclic Galois extension of degree $\ell$.

Let $d=\deg f$. Since $\deg g=\ell$ and $d<\ell$, the place $P_\infty$ of $L$ lies above the zero place $P_0$ of $K$, with $e(P_\infty\mid P_0)=\ell-d$. Since $L/K$ is Galois of prime degree $\ell$, every ramification index is either $1$ or $\ell$. Hence $\ell-d\in\{1,\ell\}$, and therefore $d\in\{0,\ell-1\}$.

If $d=0$, then $f\in\F_q^\times$, so $1/h=g/f$ is a polynomial of degree $\ell$. Hence $h$ is equivalent to a polynomial.

Now suppose that $d=\ell-1$. Then $e(P_\infty\mid P_0)=1$, so $P_0$ is unramified. After extending constants to $\overline{\F}_q$, we obtain a tame cyclic extension $\overline{\F}_q(x)/\overline{\F}_q(t)$ of degree $\ell$. Let $r$ be the number of geometric branch points. Since both function fields have genus zero and every ramified point is totally ramified, the Riemann--Hurwitz formula gives
\[
-2=-2\ell+r(\ell-1),
\]
hence $r=2$.

Thus the geometric branch locus consists of exactly two points. Since the cover is defined over $\F_q$, this two-element set is stable under the $q$-power Frobenius, so both branch points are defined over $\F_{q^2}$. Since $P_0$ is unramified, neither branch value is $0$. Choose a finite branch value $b\in\F_{q^2}^\times$. There is a unique point above $b$, and it is totally ramified. Since this point is also defined over $\F_{q^2}$ and is finite, it corresponds to some $a\in\F_{q^2}$.

Therefore $f(X)-bg(X)$ has $a$ as its unique root with multiplicity $\ell$. Since $g$ is monic and $\deg f=\ell-1$, its leading coefficient is $-b$, and hence
\[
f(X)-bg(X)=-b(X-a)^\ell.
\]
If $b\in\F_q$, then the unique point above $b$ is fixed by Frobenius, so $a\in\F_q$.
\end{proof}

\begin{Remark}
\label{rem:cyclic-prime}
Consider the non-polynomial case of \cref{prop:cyclic-prime}, so $\deg f=\ell-1$ and $\deg g=\ell$. Up to postcomposition by a suitable element of $\operatorname{PGL}_2(\F_q)$ and scaling, we may assume that $f$ is monic and that the coefficient of $X^{\ell-1}$ in $g$ is zero.

By \cref{prop:cyclic-prime}, there exist $a,b\in\F_{q^2}$, with $b\neq0$, such that
\[
f(X)-bg(X)=-b(X-a)^\ell.
\]
Comparing the coefficients of $X^{\ell-1}$ gives $1=\ell ab$. In particular, if $\chara(\F_q)=2$, then $\ell$ is odd, and hence $a=b^{-1}$.

If the corresponding geometric branch values are individually defined over $\F_q$, then $a,b\in\F_q$. Otherwise, they form a Frobenius-conjugate pair over $\F_{q^2}$.
\end{Remark}

Having established the cyclic factorization behavior, we now constrain the extension degree $\ell = [\F_q(x) : \F_q(t)]$. Since the Galois group of the extension injects into the automorphism group of the rational function field, the geometry of the projective line imposes a strict divisibility condition on $\ell$.

\begin{Lemma}
\label{Lemma l}
Let $f(X), g(X) \in \F_q[X]$ be coprime polynomials over a finite field $\F_q$, and define the nonconstant rational function $t = \frac{f(x)}{g(x)} \in \F_q(x)$. Let $\ell = [\F_q(x) : \F_q(t)]$ be the degree of the corresponding function field extension. Assume that the extension $\F_q(x)/\F_q(t)$ is Galois.

Then:
\begin{enumerate}
    \item Every $\sigma \in \Gal(\F_q(x)/\F_q(t))$ induces a Möbius transformation on $\mathbb{P}^1$, i.e., there exists $h(x) \in \mathrm{PGL}_2(\F_q)$ such that
    \[
    \sigma(x) = h(x) = \frac{a x + b}{c x + d}, \quad ad - bc \neq 0.
    \]
    \item In particular, the degree $\ell$ divides the order of the group $\mathrm{PGL}_2(\F_q)$, and therefore
    \[
    \ell \mid q (q - 1)(q + 1).
    \]
\end{enumerate}
\end{Lemma}

\begin{proof}
The automorphism group of $\F_q(x)$ over $\F_q$, denoted $\Aut_{\F_q}(\F_q(x))$, is known to be:
\[
\Aut_{\F_q}(\F_q(x)) = \left\{ x \mapsto \frac{a x + b}{c x + d} : a,b,c,d \in \F_q,\ ad - bc \neq 0 \right\} \cong \mathrm{PGL}_2(\F_q).
\]
Since every element of $\Gal(\F_q(x)/\F_q(t))$ fixes $\F_q$, we have
\[
\Gal(\F_q(x)/\F_q(t))
\leq
\Aut_{\F_q}(\F_q(x))
\cong\mathrm{PGL}_2(\F_q).
\]
This proves point (1).

\medskip

Point (2) follows immediately: Since $\F_q(x)/\F_q(t)$ is Galois, its order 
\[
\abs*{\Gal(\F_q(x)/\F_q(t))} = [\F_q(x):\F_q(t)] = \ell
\]
must divide the order of $\mathrm{PGL}_2(\F_q)$, which is:
\[
\abs*{\mathrm{PGL}_2(\F_q)} = q(q^2 - 1) = q(q - 1)(q + 1).
\]
Thus, $\ell \mid q(q - 1)(q + 1)$, as claimed.
\end{proof}

With the divisibility constraints of \cref{Lemma l} in place, we can now state our main classification result. The cases of degree $2$, and of degree $4$ over fields with more than two elements, are already attained by known polynomial constructions. The following theorem determines the minimal Galois group sizes $\cG_n(q)$ for all remaining cases up to degree $6$ in characteristic $2$, utilizing both theoretical constraints and explicit verification via computational algebra systems. Together with the known polynomial cases, this completes the low-degree classification.

\begin{Thm}
\label{thm:main}
    For $m \geq 1$,
\[
\cG_n(q)= \begin{cases}
    3 & \text{ if } n = 3, q=2^m \\
    8 & \text{ if } n = 4, q=2 \\
    5 & \text{ if } n = 5, q=2^{2m} \\
    10 & \text{ if } n = 5, q=2^{2m-1} \\
    6 & \text{ if } n = 6, q=2^m 
    \end{cases}
\]
\end{Thm}

\begin{proof}
Let $h=f/g\in\F_q(x)$ be a separable rational function of degree $n$. Since $G(h)$ acts transitively on the $n$ roots of $f(X)-tg(X)$, we have
\[
n \mid \abs*{G(h)} \leq \abs*{A(h)}.
\] 
Throughout this proof, the exact sizes of the arithmetic Galois groups for the explicit rational functions provided were verified computationally using the \texttt{MAGMA} algebra system.

For $n = 3$, consider $f(x) = x^2 + x$ and $g(x) = x^3 + x + 1$ in $\F_2[x]$. We have $\abs*{A(h)} = 3$. Since $\F_2 \subset \F_{2^m}$ for all $m \geq 1$, the same pair $(f, g)$ yields $\abs*{A(h)} = \abs*{G(h)} = 3$ over all such extensions. Hence, $\cG_3(2^m) = 3$ for all $m \geq 1$.

For $n=4$ and $q=2$, suppose that $\cG_4(2)=4$. Then a rational function attaining this value would satisfy $\abs*{A(h)}=\abs*{G(h)}=4$, and hence $\F_2(x)/\F_2(h)$ would be Galois. By \cref{Lemma l}, this would imply
\[
4 \mid \abs*{\mathrm{PGL}_2(\F_2)} = 6,
\]
which is impossible. Since $4 \mid \abs*{G(h)}$, it follows that $\cG_4(2) \geq 8$. 
Consider the functions
\[
f(x) = x^2, \quad g(x) = x^4 + x^3 + x + 1 \in \F_2[x].
\]
For this pair, we get $\abs*{A(h)} = 8 = \abs*{\operatorname{Gal}((f - t g)/\F_4(t))}$. 
This implies that the constant field $\F_4$ is not contained in the splitting field $M$ of $f - t g$ over $\F_2(t)$, because the Galois group does not shrink after extending the field of constants to $\F_4$. Since $[M \cap \overline{\F}_2 : \F_2] = \abs*{A(h)}/\abs*{G(h)}$, we conclude that $\cG_4(2) = 8$.

For $n = 5$, let $\F_4 = \F_2(\alpha)$, where $\alpha^2+\alpha+1=0$. We find $\abs*{A(h)} = 5$ when
\[
f(x) = x^4 + x, \quad g(x) = x^5 + \alpha x + \alpha \in \F_4[x].
\]
The same holds for every extension field containing $\F_4$, and therefore $\cG_5(2^{2m}) = 5$.

However, using \cref{Lemma l}, observe that
\[
5 \nmid (2^{2m-1} - 1) \cdot 2^{2m-1} \cdot (2^{2m-1} + 1),
\]
so $\cG_5(2^{2m-1}) > 5$ for all $m \geq 1$. Since $5 \mid \abs*{G(h)}$, we obtain $\cG_5(2^{2m-1}) \geq 10$. Consider the functions
\[
f(x) = x^4 + x^2, \quad g(x) = x^5 + x^3 + 1 \in \F_2[x].
\]
For this pair, we get $\abs*{A(h)} = \abs*{G(h)} = 10$, hence $\cG_5(2^{2m-1}) = 10$.

Finally, for $n=6$, consider
\[
h(x)=\frac{x^4+x^2}{x^6+x^5+x^3+x+1}\in\F_2(x).
\]
A direct computation shows that $h$ is invariant under the transformations $x \mapsto x+1$ and $x \mapsto \frac{1}{x}.$ These transformations generate $\mathrm{PGL}_2(\F_2)$, which has order $6$. Since $\deg h=6$, the extension $\F_2(x)/\F_2(h)$ is Galois with Galois group $\mathrm{PGL}_2(\F_2)$. Therefore $A(h)=G(h)\cong\mathrm{PGL}_2(\F_2),$ and the same remains true over every $\F_{2^m}$. Consequently, $\cG_6(2^m)=6.$
\end{proof}

\section{Construction of Optimal LRCs Using Good Rational Functions}
\label{Sec:Construction}
The minimal Galois groups classified in the previous section control the number of totally split places in the corresponding function field extensions. We now translate these arithmetic properties into the construction of optimal LRCs. By utilizing rational functions attaining the minimal Galois group sizes determined in \cref{thm:main}, we can generate evaluation sets of maximal size on the affine line, outperforming the lengths achievable by generic polynomial constructions.

We construct an evaluation code using a rational function $\phi$. This construction follows the rational-map framework of Munuera and Ten\'orio~\cite{MUNUERA}, extending the classical Tamo--Barg construction to rational functions. Let $\A(\F)$ be the affine line over $\F$ and $A_1,A_2,\ldots,A_\ell \subset \A(\F)$ be $\ell>1$ pairwise disjoint subsets, each of cardinality $r+1$. A key ingredient of the construction is a rational function $\phi(x)\in \F(x)$ with no poles in $A = A_1 \cup \dots \cup A_\ell$ and satisfying:
    \begin{itemize}
        \item The degree of $\phi$ is $r+1$.
        \item $\phi$ is constant on each $A_i$, $i=1,\ldots,\ell$.
    \end{itemize}
A rational function satisfying these conditions is good in the sense defined above. In particular, the sets $A_i$ may be chosen as distinct finite totally split fibers of $\phi$. The construction below depends on the existence of such good rational functions.

Up to postcomposition by an element of $\mathrm{PGL}_2(\F)$, we may assume that $\phi(\infty)=\infty$. Thus, writing $\phi=f/g$ in reduced form, we have $\deg f>\deg g$, and every totally split fiber above a finite value is contained entirely in $\A(\F)$.

\begin{constr}
    Set $n = \ell(r+1)$ and fix a positive integer $k$ such that $r \mid k$ and $k + \frac{k}{r} \leq n$. Consider the $\F$-linear space $V \subset \F(x)$ defined by 
\[
V = \left\{ \sum_{i = 0} ^{r-1} \sum_{j=0} ^{\frac{k}{r}-1} a_{ij} \phi(x)^j x^i : a_{ij}\in\F\right\}.
\] 
Since $[\F(x):\F(\phi)]=r+1$, the functions appearing in the definition of $V$ are linearly independent over $\F$, and hence $\dim(V)=k$. The code $\cC$ is obtained by evaluating the functions in $V$ at the points of $A$ via $\operatorname{ev}_A:V\longrightarrow\F^n$. Since $\phi$ is constant on each $A_s$, the restriction of every $v\in V$ to $A_s$ is a polynomial in $x$ of degree at most $r-1$. Hence, any erased coordinate can be recovered from the remaining $r$ coordinates corresponding to the points of $A_s$ by interpolation. Since $\deg(\phi)=r+1$, every $v\in V$ satisfies
\[
\deg(v)\leq k+\frac{k}{r}-2<n.
\]
Thus, $\cC$ is a linear $(n,k,r)$ LRC. Since a nonzero rational function of degree $D$ has at most $D$ zeros, the minimum distance satisfies
\[
 d(\cC) \geq n-\left(k+\frac{k}{r}-2\right) = n-k-\frac{k}{r}+2. 
\] 
Since $r\mid k$, the Singleton-type bound gives $d(\cC)\leq n-k-\frac{k}{r}+2$. Thus equality holds, and $\cC$ is optimal.
\end{constr}

\section{On the Number of Totally Split Places}
\label{sec:LRS}
To determine the maximum length of the LRCs constructed in \cref{Sec:Construction}, we need to determine the number of totally split places in the extension $\F_q(x)/\F_q(t)$. We approach this in two steps. First, we establish asymptotic bounds using the Hasse-Weil framework together with
estimates for the genera of the splitting fields for our characteristic $2$ functions. Second, we move beyond these asymptotic estimates by proving that the exact number of totally split places across field extensions satisfies a linear recurring sequence, allowing for exact parameter computation without relying on error bounds.

\subsection{Estimate of the Genus}

Let $h=f/g\in\F_q(x)$ be a separable rational function of degree $n\geq2$ satisfying $\abs*{A(h)}=\abs*{G(h)}=\cG_n(q)$, and let $M$ be the splitting field of $f(X)-tg(X)$ over $\F_q(t)$. Denote by $g_M$ the genus of $M$ and let $\alpha, \beta \in M$ be the roots of $f-gt$. Furthermore, assume that $F_1 = \F_q (\alpha)$ and $F_2 = \F_q (\beta)$ where $g_{F_1}, g_{F_2}$ are the genera of $F_1,F_2$, respectively. We first obtain genus estimates for the cases $\cG_n(q)=n$ and $\cG_n(q)=2n$.

\begin{Lemma}
\label{lemma1}
      If $\cG_n(q)=n$, then $g_M=0$.
\end{Lemma}
\begin{proof}
  It follows from \cite[Proposition 1.6.3]{Stichtenoth}.
\end{proof}

\begin{Lemma}
\label{lemma2}
    If $\cG_n(q)=2n$, then $g_M\leq1$.
\end{Lemma}
\begin{proof}
    We can choose $\alpha,\beta$ such that $F_1\neq F_2$. By \cite[Proposition 1.6.3]{Stichtenoth}, $g_{F_1}=g_{F_2}=0$. It is easy to check that $M=\F_q(\alpha, \beta)=F_1F_2$, therefore by Castelnuovo's Inequality \eqref{eq:castelnuovo} $g_M\leq 1$.
\end{proof}

The upper bound established in \cref{lemma2} is attained in the specific cases needed below.

\begin{Lemma}
\label{lem:special-genera}
In the cases $(n,q)=(4,2)$ and $(n,q)=(5,2^{2m-1})$, where $m\geq1$, we have $g_M=1$.
\end{Lemma}

\begin{proof}
By \cref{lemma2}, we have $g_M\leq1$. Suppose that $g_M=0$. Since $\abs*{A(h)}=\abs*{G(h)}$, the full constant field of $M$ is $\F_q$. Thus, \eqref{eq:rational-place-count} shows that $M$ has an $\F_q$-rational place, and hence $M\cong\F_q(u)$ by~\cite[Proposition~1.6.3]{Stichtenoth}. Therefore,
\[
A(h)
=
\Gal(M/\F_q(t))
\leq
\Aut_{\F_q}(M)
\cong
\mathrm{PGL}_2(\F_q),
\]
so $\abs*{A(h)}$ must divide $q(q-1)(q+1)$. This is impossible in the degree-$4$ case, since $8\nmid6$, and in the degree-$5$ case, since
\[
5\nmid
2^{2m-1}\bigl(2^{2m-1}-1\bigr)\bigl(2^{2m-1}+1\bigr).
\]
Consequently, $g_M\neq0$, and therefore $g_M=1$.
\end{proof}

\subsection{Estimates for Totally Split Places}

Recall that $T_{\mathrm{split}}^1(h)$ denotes the set of rational places of $\F_q(t)$ that split completely in $\F_q(x)/\F_q(t)$, where $t=h(x)$.

\begin{Prop}[Corollary 4.7,~\cite{Beyond_polynomials}]
\label{prop:HW-split}
Let $h=f/g$ be a rational function with $A(h)=G(h)$. Let $t=h(x)$, and let $g_M$ be the genus of the splitting field $M$ of $f(X)-tg(X)$ over $\F_q(t)$. Then
\[
\left\lceil
\frac{q+1-2g_M\sqrt{q}}{\abs*{G(h)}}
-\frac{\abs*{R_h^1}}{2}
\right\rceil
\leq
\abs*{T_{\mathrm{split}}^1(h)}
\leq
\left\lfloor
\frac{q+1+2g_M\sqrt{q}}{\abs*{G(h)}}
\right\rfloor,
\]
where $R_h^1$ denotes the set of rational places of $\F_q(t)$ that ramify in $M/\F_q(t)$.
\end{Prop}

\begin{Prop}
\label{prop:ramified-place-bound}
Let $h=f/g\in\F_q(x)$ be a separable rational function satisfying $A(h)=G(h)$, and let $M$ be the splitting field of $f(X)-tg(X)$ over $\F_q(t)$. Set
\[
N=[M:\F_q(t)]=\abs*{G(h)}.
\]
If $R_h^1$ denotes the set of rational places of $\F_q(t)$ that ramify in $M/\F_q(t)$, then
\[
\abs*{R_h^1}
\leq
\left\lfloor
4+\frac{4g_M-4}{N}
\right\rfloor.
\]
\end{Prop}

\begin{proof}
Since $M$ is the splitting field of the separable polynomial $f(X)-tg(X)$, the extension $M/\F_q(t)$ is Galois. Fix $P\in R_h^1$, and let $e_P$, $f_P$, and $s_P$ denote, respectively, the common ramification index, the common relative degree, and the number of places of $M$ lying above $P$. By \eqref{eq:fundamental-equality},
\[
s_Pe_Pf_P=N.
\]
Moreover, the different exponent satisfies $d(Q\mid P)\geq e_P-1$ for every $Q\mid P$. Since $\deg P=1$, we obtain
\[
\sum_{Q\mid P}d(Q\mid P)\deg Q
\geq s_P(e_P-1)f_P
=N\left(1-\frac{1}{e_P}\right)
\geq\frac{N}{2},
\]
because $e_P\geq2$. Summing over all $P\in R_h^1$ gives
\[
\deg\bigl(\operatorname{Diff}(M/\F_q(t))\bigr)
\geq
\frac{N}{2}\abs*{R_h^1}.
\]
On the other hand, the Riemann--Hurwitz formula gives
\[
\deg\bigl(\operatorname{Diff}(M/\F_q(t))\bigr)
=2g_M-2+2N.
\]
Therefore,
\[
\abs*{R_h^1}
\leq
4+\frac{4g_M-4}{N}.
\]
Since $\abs*{R_h^1}$ is an integer, the stated bound follows.
\end{proof}

Applying \cref{prop:HW-split} and \ref{prop:ramified-place-bound}, together with the values of $\cG_n(q)$ and $g_M$ obtained above, yields the following lower bounds in \cref{tab:split-place-bounds}. Whenever the resulting expression is negative, we replace it by the trivial lower bound $0$.

\begin{table}[htbp]
    \centering
    \renewcommand{\arraystretch}{1.25}
    \begin{tabularx}{\linewidth}{c c c c >{\centering\arraybackslash}X}
    \toprule
    $n$ & $q=2^m$ & $\cG_n(q)$ & $g_M$ &
    Lower bound for $\abs*{T_{\mathrm{split}}^1(h)}$ \\
    \midrule
    2 & $m\geq1$ & 2 & 0 &
    $\dfrac{q}{2}$ \\
    \midrule
    3 & $m\geq1$ & 3 & 0 &
    $\left\lceil\dfrac{q-2}{3}\right\rceil$ \\
    \midrule
    4 & $m\geq2$ & 4 & 0 &
    $\dfrac{q-4}{4}$ \\
      & $m=1$ & 8 & 1 &
    $0$ \\
    \midrule
    5 & $m$ even & 5 & 0 &
    $\left\lceil\dfrac{2q-13}{10}\right\rceil$ \\
      & $m$ odd & 10 & 1 &
    $\max\left\{0,
    \left\lceil\dfrac{q-2\sqrt q-19}{10}\right\rceil
    \right\}$ \\
    \midrule
    6 & $m\geq1$ & 6 & 0 &
    $\max\left\{0,
    \left\lceil\dfrac{q-8}{6}\right\rceil
    \right\}$ \\
    \bottomrule
    \end{tabularx}
    \caption{Lower bounds for the number of totally split rational places for rational functions of degree at most $6$.}
    \label{tab:split-place-bounds}
\end{table}

\subsection{On the Exact Number of Totally Split Places}
Although the Hasse--Weil bounds above provide estimates for the number of totally split places, the associated $\mathcal{O}(\sqrt{q})$ error term does not determine the exact number over a given extension field. We now show that the sequence counting the totally split places and the finite totally split fibers over the constant field extensions $\F_{q^i}$ are linear recurring sequences. In particular, this determines exactly the maximum length (and dimension) obtainable from the corresponding LRC construction over each extension field. 

Let $h = f/g \in \F_q(x)$ be a rational function, where $f,g$ are coprime, with $h(\infty) = \infty $, and let $M$ be the splitting field of the separable polynomial $f(X) - t g(X)$ over $ K = \F_q(t)$. Suppose that $h$ has at least one totally split rational place Then the full constant field of $M$ is $\F_q$; see, for example, \cite{bartoli2022algebraic}. 

For every $i\geq1$, set
\[
K_i=\F_{q^i}(t),\qquad M_i=M\F_{q^i},\qquad N=[M:K].
\]
Since $M/K$ is Galois and $\F_q$ is the full constant field of $M$, the extension $M_i/K_i$ is Galois of degree $N$. Let
\[
s_i=\#\left\{P\in\cP_{K_i}^1:P\text{ splits completely in }M_i/K_i\right\}
\]
and
\[
a_i=\#\left\{t_0\in\F_{q^i}:f(X)-t_0g(X)\text{ splits into }n\text{ distinct linear factors over }\F_{q^i}\right\}.
\]
Since $h(\infty)=\infty$, the number $a_i$ counts precisely the finite places among the places counted by $s_i$. Therefore, $ a_i=s_i-\epsilon_i,$ where
\[
\epsilon_i=
\begin{cases}
1,&\text{if the infinite place $P_{\infty,i}$ of $K_i$ splits completely in }M_i/K_i,\\
0,&\text{otherwise}.
\end{cases}
\]
To account for the rational places of $M_i$ lying above ramified places of $K_i$, set
\[
c_i=\#\left\{Q\in\cP_{M_i}^1:Q\text{ is ramified in }M_i/K_i\right\}.
\]

\begin{Lemma}
\label{lem:periodic-corrections}
The sequences $(c_i)_{i\geq1}$ and $(\epsilon_i)_{i\geq1}$ are periodic.
\end{Lemma}

\begin{proof}
Let $Q_1,\ldots,Q_r \in \cP_M$ be the ramified places of $M/K$, and set $\lambda_j=\deg Q_j$. Constant field extensions preserve ramification, and a place of degree $\lambda_j$ gives rise to $\lambda_j$ rational places over $\F_{q^i}$ if $\lambda_j\mid i$ and to no rational places otherwise~\cite[Theorem~3.6.3(a) and Lemma~5.1.9(d)]{Stichtenoth}. Therefore,
\[
c_i=\sum_{\substack{1\leq j\leq r\\\lambda_j\mid i}}\lambda_j.
\]
Consequently, if $L=\operatorname{lcm}(\lambda_1,\ldots,\lambda_r)$, then $c_{i+L}=c_i$ for every $i\geq1$.

If $P_{\infty,1}$ is ramified in $M/K$, then it remains ramified under every constant field extension, and hence $\epsilon_i=0$ for every $i\geq1$. Otherwise, let $f_\infty$ be the common relative degree of the places of $M$ above $P_{\infty,1}$. Since $P_{\infty, 1}$ is rational and unramified, there are $N/f_\infty$ such places, each of degree $f_\infty$, and again Lemma~5.1.9(d) of~\cite{Stichtenoth} shows that $P_{\infty,i}$ splits completely in $M_i/K_i$ if and only if $f_\infty\mid i$. Thus, $\epsilon_{i+f_\infty}=\epsilon_i$ for every $i\geq1$.
\end{proof}

\begin{Thm}
\label{thm:split-place-lrs}
The sequences $(s_i)_{i\geq1}$ and $(a_i)_{i\geq1}$ are linear recurring sequences.
\end{Thm}

\begin{proof}
Set $\nu_i=\#\cP_{M_i}^1$. By~\cref{eq:rational-place-count}, $(\nu_i)_{i\geq1}$ is a linear recurring sequence. Let $Q\in\cP_{M_i}^1$ be a place not counted by $c_i$, and let $P=Q\cap K_i$. Since
\[
1=\deg Q=f(Q\mid P)\deg P,
\]
we have $\deg P=f(Q\mid P)=1$. Moreover, $e(Q\mid P)=1$. Since $M_i/K_i$ is Galois, the fundamental equality shows that $P$ splits completely in $M_i/K_i$. Conversely, every rational place of $K_i$ that splits completely in $M_i/K_i$ has exactly $N$ rational places of $M_i$ lying above it. Therefore,
\[
\nu_i=Ns_i+c_i.
\]
By~\cref{lem:periodic-corrections}, $(c_i)_{i\geq1}$ is periodic and therefore is a linear recurring sequence. It follows that $(s_i)_{i\geq1}$ is a linear recurring sequence. Finally, since $a_i=s_i-\epsilon_i$ and $(\epsilon_i)_{i\geq1}$ is periodic by~\cref{lem:periodic-corrections}, $(a_i)_{i\geq1}$ is also a linear recurring sequence.
\end{proof}

\section*{Acknowledgements}

The authors thank Giacomo Micheli for his guidance and helpful discussions during the early stages of this work.

\noindent The authors used ChatGPT (OpenAI) to assist with drafting and revising the manuscript. Authors independently reviewed and verified the mathematical content, revised the text as needed, and take full responsibility for the content of the paper.

\bibliography{reference}
\bibliographystyle{abbrv}

\end{document}